%% file: nanolett.tex
\documentclass[journal=jacsat,manuscript=article]{achemso}

\usepackage[version=3]{mhchem} 
\usepackage{amsmath,bm}
\usepackage{amssymb,amsthm,braket}
\usepackage{color}
\usepackage{pifont}

\author{Seoung-Hun Kang}
\affiliation{Department of Physics and
             Research Institute for Basic Sciences,
             Kyung Hee University, Seoul, 02447, Korea}
\alsoaffiliation{Korea Institute for Advanced Study, Seoul, 02455, Korea}
\author{Jejune Park}
\altaffiliation{Current address: IMEP-LaHC, Grenoble INP,
                3 parvis Louis N\'eel, 38016 Grenoble, France}
\affiliation{Department of Physics and
             Research Institute for Basic Sciences,
             Kyung Hee University, Seoul, 02447, Korea}
\author{Sungjong Woo}
\email{sjwoo@ibs.re.kr}
\affiliation{Korea Institute for Advanced Study, Seoul, 02455, Korea}
\alsoaffiliation{Institute for Basic Sciences, Daejeon, 34126, Korea}
\author{Young-Kyun Kwon}
\email{ykkwon@khu.ac.kr}
\affiliation{Department of Physics and
             Research Institute for Basic Sciences,
             Kyung Hee University, Seoul, 02447, Korea}
\alsoaffiliation{Korea Institute for Advanced Study, Seoul, 02455, Korea}

\title{Symmetry-Protected Degeneracies in the Electronic Band Structure of Oxidized Black Phosphorous}


\begin{document}

%

\begin{abstract}
\input{abstract}
\end{abstract}

\input{maintext}

\begin{acknowledgement}
\input{acknowledgements}
\end{acknowledgement}

%

\input{My_collection.bbl}

\end{document}

%% file: abstract.tex
We explore the oxidation of a single layer of black phosphorous using \textit{ab initio} density functional theory calculation. We search for the equilibrium structures of phosphorene oxides, PO$_x$ with various oxygen concentrations $x$ ($0{\le}x{\le}1$). By evaluating the formation energies with diverse configurations and their vibrational properties for each of various $x$ values, we identify a series of stable oxidized structures with $x$ and confirm that the oxidation occurs naturally. We also find that oxidation makes some modes from the P-O bonds P-P bonds IR-active implying that the infrared spectra can be used to determine the degree of oxidation of phosphorene. Our electronic structure calculation reveals that the fully oxidized phosphorene (PO) has a direct band gap of 0.83~eV similar to the pristine phosphorene. Intriguingly, the PO possesses two nonsymmorphic symmetries with the inversion symmetry broken, guaranteeing a symmetry-protected band structures including the band degeneracy and four-fold degenerate Dirac points. Our results may provide a significant insight into the intriguing relations between symmetry of lattice and band topology of electrons.

%% file: maintext.tex
%
%
\section{Introduction}
\label{Introduction}

Since its synthesis and fabrication~\cite{{Li2014},{Koenig2014},{Ye2014}}, a single layer of black phosphorous or phosphorene has attracted a lot of  attention due to its inherent direct band gap. The energy gap varies from $1.9$~eV (monolayer) to $0.3$~eV (bulk) according to the number of layers~\cite{Guo2015}. Its high hole mobility and on/off ratio make phosphorene a potential candidate for future electronic devices~\cite{{Gomez2014},{Koenig2014},{Li2014},{Ye2014}}. On the other hand, it has been reported that single or few-layered phosphorene is highly reactive with air, especially, in oxygen environment~\cite{{Koenig2014},{Island2015}} so that oxidation would be inevitable during the synthesis or fabrication processes. The oxidation increases surface roughness and contact resistance resulting in reduction of carrier mobility~\cite{Koenig2014}. Another study showed that an oxidation process changes its electronic properties drastically~\cite{Island2015}. Thus it is necessary to understand the oxidation mechanism and its effects. Recent studies proposed possible equilibrium structures of oxidized phosphorene, PO$_x$, with various oxygen concentration values of $x$~\cite{{Ziletti2015},{Ziletti2015a},{Wang2015}}, suggesting possible oxidation mechanisms involving reactive dangling bonds~\cite{Ziletti2015}.

In this paper, we report our study on the structural, electronic, and vibrational properties of phosphorene oxides, PO$_x$, with $x$ varying from 0 to 1. It was found that the structural symmetry of phosphorene gets lowered in the process of oxidation. Our electronic structure calculation reveals that the band gap increases with $x$ being maximized near $x=0.4$, and then decreases. Although the oxidation breaks the inversion symmetry underlying the puckered structure of phosphorene, we found that two additional nonsymmorphic symmetries remain in certain structures of oxidized phosphorene, for instance PO with $x=1$. These nonsymmorphoic symmetries guarantee a four-fold degeneracy at the X point and two-fold degeneracies along the $\Gamma-$X and the X$-$S lines in the Brillouin zone. Such symmetry-protected features are accessible with reasonable electronic doping. Furthermore, the oxidation makes phosphorene active to infrared (IR) absorption with high frequency and the IR active modes are blue-shifted with oxidation. This suggests that the degree of oxidation can be experimentally determined using the IR spectrum. 

\section{Computational details}
\label{Computational}

To identify the equilibrium structures of phosphorene oxide and investigate their structural, electronic, and vibrational properties, we carried out first-principles calculations based on density  functional theory (DFT) using Vienna \textit{ab initio} simulation package (VASP)~\cite{{Kresse1996-1},{Kresse1996-2}}. Projector augmented wave potentials~\cite{Blochl1994} were employed to describe the valence electrons. The exchange-correlation functional is treated within the generalized gradient approximation (GGA) of Perdew, Burke,and Ernzerhof (PBE)~\cite{Perdew1996}. The cutoff energy for the plane wave basis is chosen to be 450~eV. The Brillouin zone is sampled  using $\Gamma$-centered $30{\times}30{\times}1$ grid.  We used a $2\times2$ supercell containing 16 phosphorous atoms and a certain number of oxygen atoms ranging from 0 to 16 in order to explore the oxidation process of phosphorene with various oxygen concentrations. To avoid the spurious inter-layer interaction, we introduced a vacuum region of 15~{\AA} along the $c$ axis perpendicular to the sheet. Atomic relaxations were done until the Helmann-Feynman force acting on every atom becomes smaller than 0.01~eV/{\AA}. The vibrational properties of some selected phosphorene oxides were evaluated using the harmonic approximation implemented in PHONOPY package~\cite{Togo2008}. To minimize their imaginary flexural modes, we used $6{\times}6{\times}1$ and $4{\times}4{\times}1$ supercell structures for partially and fully oxidized phosphorenes, respectively. We used a dipole approximation within the density functional perturbation theory~\cite{{Pick1970},{Kresse1995},{Baroni2001},{Karhanek2010}} to calculate their infrared intensities $I(\omega)$ as a function of frequency $\omega$ in terms of the oscillator strengths determined by the Born effective charge tensors and the displacement vectors:
\[
  I(\omega)=\sum_i\left|\sum_n\sum_j
    Z^*_{n,ij}e_{n,j}(\omega)\right|^2,
\]
where the index $n$ indicates the different atoms, $i$ and $j$ the Cartesian polarization, and $Z^*_{n,ij}$ and $e_{n,j}(\omega)$ are the Born effective charge tensor and the normalized vibrational eigenvector corresponding to $\omega$, respectively.


\section{Results and discussion}
\label{Results}

\begin{figure*}
\includegraphics[width=1.0\textwidth]{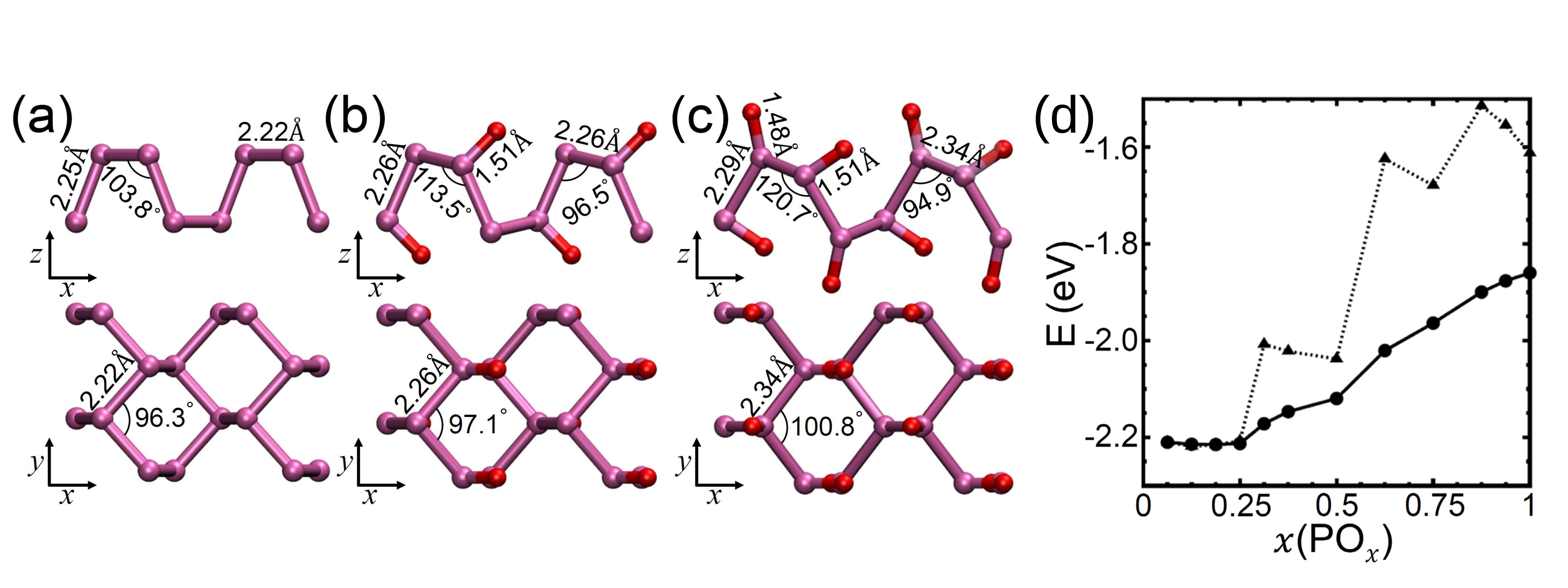}
\caption{
Side and top views of optimized structures in $2\times2\times1$ supercell for (a) pristine phosphorene and phosphorene oxides at two specific oxygen concentrations, (b) PO$_{0.5}$, and (c) PO. Some of their bond lengths and bond angles are also given in the respective configurations. The phosphorus and oxygen atoms are depicted by pink and red colors, respectively. (d) Formation energy (solid line) and energy gain (dotted line) of PO$_x$ as a function of the oxygen concentration $x$. The energy equations are given in Eqs.~(\ref{Eform}) and (\ref{Egain}) in the text.
\label{Fig1}}
\end{figure*}

We first investigated the binding process of oxygen atoms by producing PO$_x$ with $0\le x \le 1$. Various oxygen binding sites were considered on phosphorene shown in Fig.~\ref{Fig1}(a), where each phosphorous atom has $sp^3$-like bonding character with three nearest P atoms and a lone pair forming a puckered structure consisting of two P planes. Three P--P bonds can be categorized into two types, two in-plane and one inter-plane bonds. It was found that the lone pair binding site is more stable than any other binding sites with the binding energy of 2.14~eV. We further perform the oxygen binding process by increasing the oxygen concentration $x$. For each oxygen concentration, there are several configurations with different oxygen distributions. Among all such possible configurations, we were able to identify the equilibrium structure. Figure~\ref{Fig1}(b) and (c) are the equilibrium structures of PO$_x$ with $x=0.5$ and $x=1$, respectively. As oxidation proceeds, both in-plane and inter-plane P--P bonds get elongated from 2.22~{\AA} to 2.34~{\AA}, and from 2.25~{\AA} to 2.29~{\AA}, respectively, as shown in Fig.~\ref{Fig1}(a--c). All the P--O bonds in PO$_{0.5}$ are equivalent with the bond length of 1.51~{\AA}, while PO has two distinct types of P--O bonds with 1.48~{\AA} and 1.51~{\AA}. The bond angle between in-plane and inter-plane P--P bonds changes from 103.8$^\circ$ in the pristine phosphorene to either 94.9$^\circ$ or 120.7$^\circ$ in PO through 96.5$^\circ$ or 113.5$^\circ$ in PO$_{0.5}$. The bond angle between two in-plane P--P bonds gets larger under oxidation from 96.3$^\circ$ for $x=0$ to 97.1$^\circ$ and to 100.8$^\circ$ for $x=0.5$ and $x=1$ in PO$_x$. The lattice constants $a$ and $b$ along the $x$ and $y$ directions of PO$_{0.5}$ (PO) get enlarged by 2.8~\% (11.2~\%) and 2.4~\% (9.1~\%), respectively. Oxidation, thus, makes a phosphorene distorted and expanded.

To verify whether the oxidation process naturally occurs in the oxygen environment, we calculated the formation energy and the energy gain as a function of oxygen concentration shown in Fig.~\ref{Fig1}(d). The formation energy $E_f(n)$ with the number of added oxygen atoms $n$ corresponding to $x=n/16$ in the $2\times2$ supercell containing 16 P atoms is defined by 
\begin{equation}
  E_f(n)=\frac{1}{n}\left[E_\textrm{PO}(n)-\left(E_\textrm{P}+{\frac{n}{2}}E_{\textrm{O}_2}\right)\right],
  \label{Eform}
\end{equation}
where $E_\textrm{PO}(n)$ and $E_\textrm{P}$ are the total energies of PO with $n$ oxygen atoms and phosphorene in the supercell used, and $E_{{\rm O}_2}$ that of an oxygen molecule. Although the absolute values of the formation energy decrease as the oxygen concentration increases, the formation energy remains negative up to $x=1$ indicating that the oxidation is energetically preferred. Since the formation energy, however, may not guarantee successive oxidation steps, we also computed the energy gain $\Delta E(n)$ with $x=n/16$ for every oxidation step from $n-1$ to $n$, which is defined by 
\begin{align}
  \Delta E(n)
    &=E_\mathrm{PO}(n)-\left[E_\mathrm{PO}(n-1)+\frac{1}{2}E_{\mathrm{O}_2}\right] \nonumber \\
    &=nE_f(n)-(n-1)E_f(n-1).
    \label{Egain}
\end{align}
As shown in Fig.~\ref{Fig1}(d), the energy gain also remains negative verifying that the oxidation can successively progress one oxygen atom by one. 

\begin{figure}
\includegraphics[width=1.0\columnwidth]{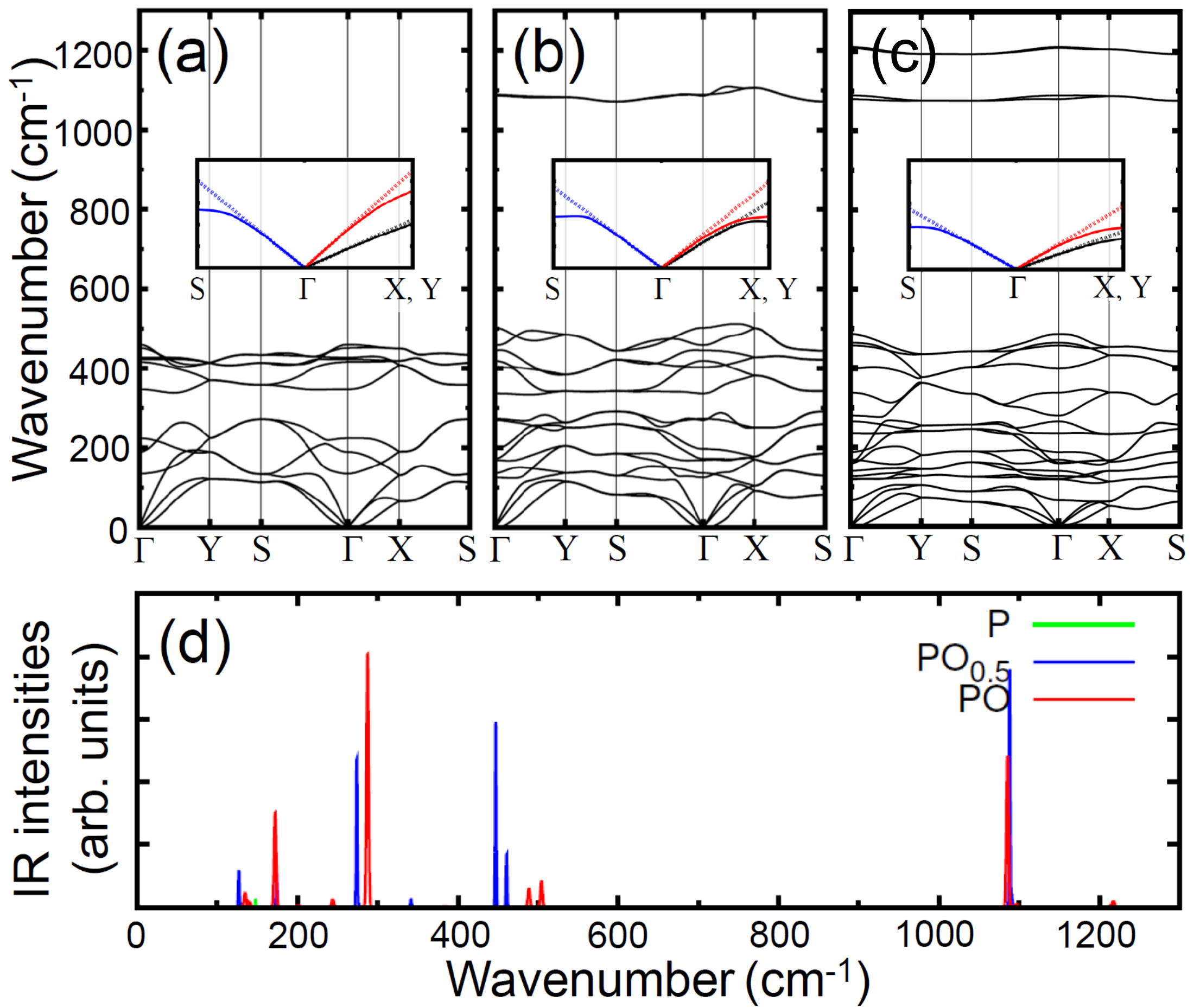}
\caption{
Phonon dispersion relations of (a) pristine phosphorene, (b) PO$_{0.5}$, and (c) PO. Each inset displays the longitudinal acoustic branches (solid lines) near the $\Gamma$ point along the three directions from the $\Gamma$ point toward the S (blue), X (black) and Y (red) points and their corresponding slopes (dotted lines) at the $\Gamma$ point indicating the speeds of sound, whose values are summarized in Table~\ref{Table}. (d) Infrared spectra of phosphorene (green) and phosphorene oxides, PO$_{0.5}$ (blue), and PO (red). 
\label{IR}}
\end{figure}

To verify the structural stability of phosphorene oxides, we computed the phonon dispersion relations of P, PO$_{0.5}$, and PO shown in Fig.~\ref{IR}(a--c). The phonon dispersion of the phosphorene and PO are in good agreement with other studies~\cite{Zhu2014,Lee2019}. Absence of negative values in the phonon frequencies indicates dynamical stability for all three configurations. All phonon frequencies of the pristine phosphorene are smaller than 500~cm$^{-1}$ as shown in Fig.~\ref{IR}(a). There are, on the other hand, a nearly-flat phonon band with a higher frequency of 1100~cm$^{-1}$ in PO$_{0.5}$ displayed in Fig.~\ref{IR}(b), corresponding to the stretching vibration of the P--O bond shown in Fig.~\ref{Fig1}(b). Figure~\ref{IR}(c) shows two phonon flat bands with higher frequencies of 1100~cm$^{-1}$ and 1200~cm$^{-1}$ indicating vibration modes from two distinct P--O bonds depicted in Fig.~\ref{Fig1}(c). The higher frequency modes corresponds to the shorter bond length.

\begin{table}
\caption{
Speeds of sound along the $\Gamma-$X, $\Gamma-$Y, and $\Gamma-$S directions of P, PO$_{0.5}$, and PO evaluated from the insets in Fig.~\ref{IR}(a--c). Values are given in the unit of km/s. 
\label{Table}}
\begin{tabular}{c|ccc}
 \hline \hline
  Sound Velocity & P & PO$_{0.5}$ & PO \\ 
 \hline
  $v_{\Gamma-\mathrm{X}}$ & 4.2 & 5.7 & 3.3\\ 
  $v_{\Gamma-\mathrm{Y}}$ & 8.4 & 7.5 & 5.4 \\ 
  $v_{\Gamma-\mathrm{S}}$ & 7.6 & 7.0 & 5.2 \\
\hline \hline
\end{tabular}
\end{table}

We further investigate the in-plane stiffness by computing sound velocities corresponding to the slopes of the longitudinal acoustic (LA) branches near $\Gamma$,~\cite{Zhu2014} as shown in the insets of Fig.~\ref{IR}(a--c). The evaluated speeds of sound are summarized in Table~\ref{Table}. It clearly shows significant anisotropy in rigidity as expected from the strong anisotropic puckered structures. For all three configurations, the speeds of sound along the $\Gamma-$Y direction are much higher than along the $\Gamma-$X direction well matched with their directional rigidity. Our results also indicate that the oxidation process usually makes the LA modes softened with an exception that the LA frequencies along the $\Gamma-$X are higher in PO$_{0.5}$ than in P. 

From the phonon calculations, we also computed the IR-active modes of the three optimized configurations of PO$_x$ with $x=0, 0.5$, and 1. Figure~\ref{IR}(d) shows our calculated IR intensities. For pristine phosphorene, there is only one weak peak at 148~cm$^{-1}$ corresponding to an out-of-plane mode. Their counter peaks for PO$_{0.5}$ and PO are observed at 127~cm$^{-1}$ and 134~cm$^{-1}$, respectively. Oxidation generated other IR-active modes related to collective motions mainly by phosphorus atom at 173~cm$^{-1}$, 274~cm$^{-1}$, 342~cm$^{-1}$, 448~cm$^{-1}$, and 462~cm$^{-1}$ for PO$_{0.5}$, and at 173~cm$^{-1}$, 243~cm$^{-1}$, 287~cm$^{-1}$, 488~cm$^{-1}$, and 506~cm$^{-1}$ for PO. It turns out that the P--O bond stretching modes observed in Fig.~\ref{IR}(b) and (c) are also IR-active around 1100~cm$^{-1}$ for PO$_{0.5}$ and around 1100~cm$^{-1}$ and 1200~cm$^{-1}$ for PO. We suggest that IR measurement could possibly be used to determine the degree of oxidation of phosphorene in the experiments.

\begin{figure}
\includegraphics[width=1.0\columnwidth]{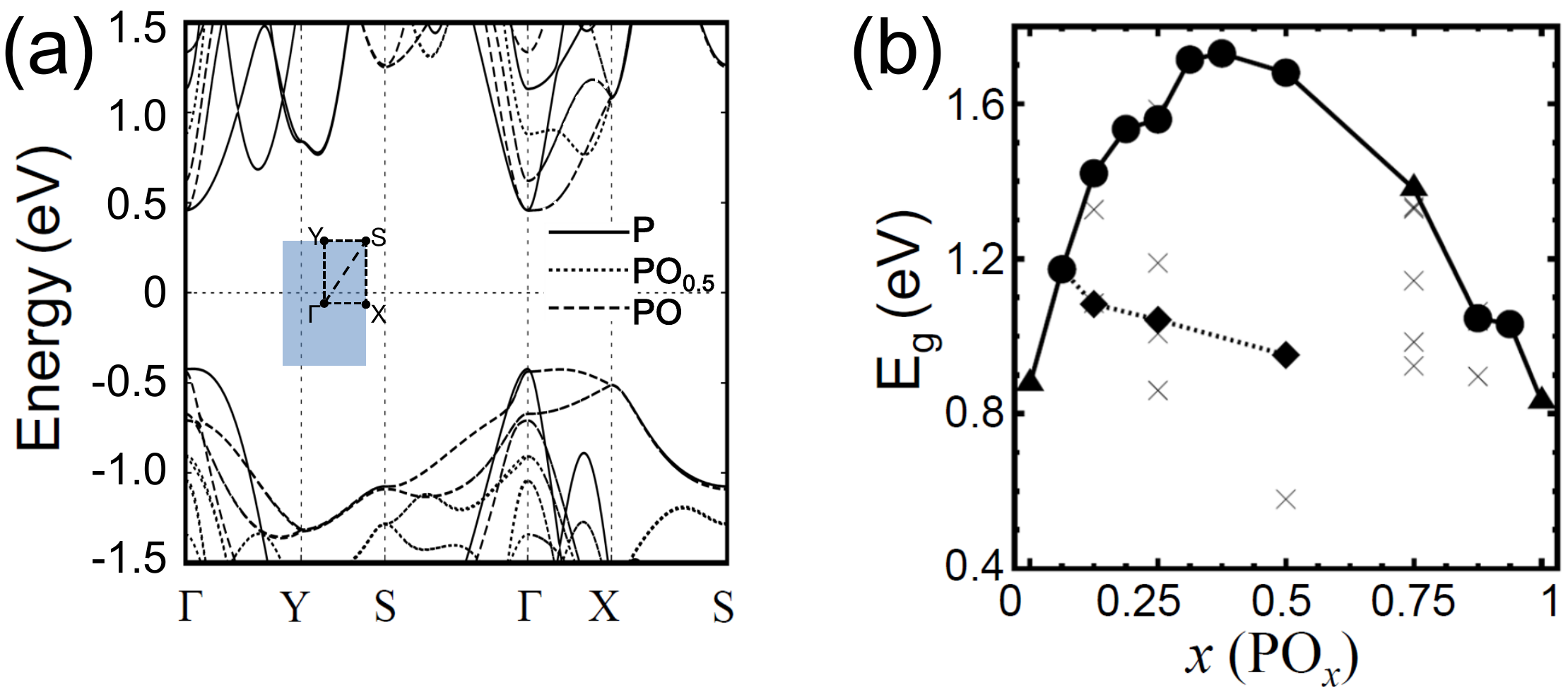}
\caption{
Electronic band structures of (a) pristine phosphorene (solid lines), PO$_{0.5}$ (dotted lines), and PO (dashed lines). The inset shows the first Brillouin zone with the special points and lines. (b) The trend of energy gap $E_g$ of phosphorene oxides PO$_x$ with $x$, the oxygen composition. Data points connected with the solid lines correspond to $E_g$ of the most stable PO$_x$ for given $x$. The \ding{108} (\ding{115}) symbol indicates an indirect (a direct) band gap. The \ding{117} symbols connected with the dotted lines represent $E_g$ of PO$_x$ with oxygen adsorbed on only one surface side. The energy band gaps of other less stable configurations are marked with the \ding{53} symbols for different $x$ values. 
\label{Fig2}}
\end{figure}

Next, we looked into the electronic band structures of various configurations of PO$_x$. As shown in Fig.~\ref{Fig2}(a), the pristine phosphorene and PO are semiconductors with direct band gap of 0.88 and 0.83~eV, respectively, observed at the $\Gamma$ point. On the other hand, PO$_{0.5}$ exhibits much wider band gap ($\sim1.68$~eV) than the pristine phosphorene. Moreover the band gap is indirect since the conduction band minimum moves from the $\Gamma$ toward the X point, while the valence band maximum stays at $\Gamma$ as shown in Fig.~\ref{Fig2}(a). Figure~\ref{Fig2}(b) shows the trend of the band gap of PO$_x$ with $x$. It tends to increase with $x$ up to $x\simeq0.4$ and then to decrease with $x$. The figure also displays the band gaps of various less stable configurations, such as those with oxygen atoms adsorbed on only one surface side.  

\begin{figure}
\includegraphics[width=1.0\columnwidth]{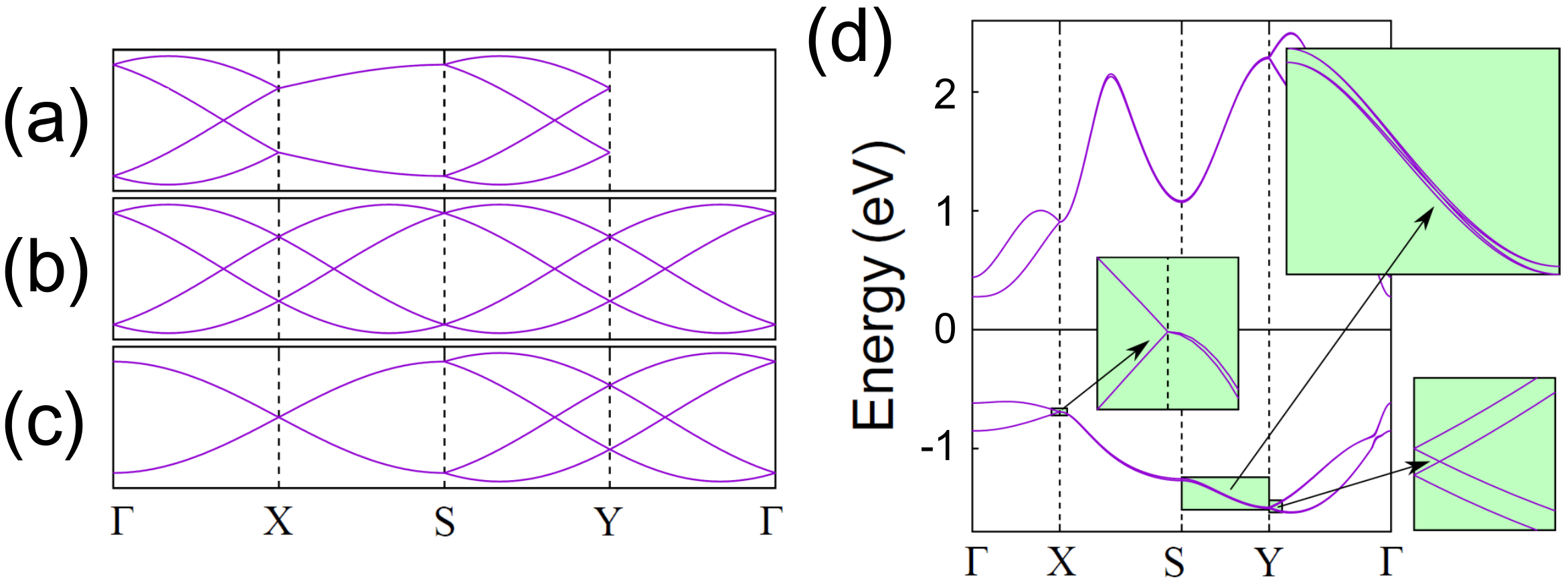}
\caption{
Schematic band structures showing the essential level crossings, partner exchange, and symmetry protected degeneracies due to (a) $\mathcal{S}=C_{2x}T_{a\hat{x}/2}$, (b) $\mathcal{G}=M_zT_{(a\hat{x}+b\hat{y})/2}$, and (c) combined symmetries. (d) Electronic band structure of real PO replotted from Fig.~\ref{Fig2}(a), containing only eight energy bands with four below and four above the Fermi level. It exhibits all the symmetry-protected features in (c) indicating the same topological symmetry as in (c). Insets are added in order to make the detailed features visible. The same features can also be observed in the four conduction bands, although it appears as if the bands were degenerate along the X$-$S and S$-$Y lines.
\label{Fig3}}
\end{figure}

The electronic structures can be characterized further by symmetry. Phosphorene possesses the inversion symmetry guaranteeing that its band lines are all doubly degenerate. Oxidation process usually reduces symmetries, such as the inversion symmetry breaking, and thus lifts some degeneracies in PO$_{0.5}$ and PO. It is, however, worth noting that there are two nonsymmorphic symmetries remaining in the optimized structures of both PO$_{0.5}$ and PO shown in Fig.~\ref{Fig1}(b) and (c).

The two nonsymmorphic symmetries underlying in PO$_x$, which are characterized by the $\mathcal{S}$crew axis and $\mathcal{G}$lide plane, can be decomposed into products of point and translation group operations as 
$\mathcal{S}=C_{2x}T_{a\hat{x}/2}$ and $\mathcal{G}=M_zT_{(a\hat{x}+b\hat{y})/2}$.
Here, $C_{2x}$ and $M_z$ are the two-fold rotation around the $x$ axis and the mirror operation about a plane perpendicular to the $z$ axis, respectively. 
$T_\mathbf{R}$ is the translation by the amount of $\mathbf{R}$ and $a$ and $b$ are the lattice constants along the $x$ and $y$ axes.
The representation of $T_\mathbf{R}$ for a Bloch state with $\mathbf{k}$ is $T_\mathbf{r}=e^{i\mathbf{k}\cdot\mathbf{R}}$ so that 
\[
  \mathcal{S}=C_{2x}e^{ik_xa/2}\quad\mbox{and}\quad \mathcal{G}=M_ze^{i(k_xa+k_yb)/2}.
\]
Furthermore, for a spin-half electron, the spinor representations of $C_{2x}$ and $M_z$ are given by
\[
  C_{2x}=i\sigma_x{\otimes}R_x(\pi)\quad\mbox{and}\quad M_z=i\sigma_z\otimes R_z(\pi)P,
\]
where, $\sigma_i$ is the $i$-th component of Pauli spin matrices; $R_i(\phi)$ and $P$ are respectively a real space rotation around the $i$-th axis by an angle $\phi$ and an inversion $\mathbf{r}\rightarrow-\mathbf{r}$. Existence of these two nonsymmorphic symmetries protect certain degeneracies and band crossing features even in phosphorene oxides without inversion symmetry. Such symmetry protected features are doubly-degenerate bands along the $\Gamma-$X and X$-$S lines and a four-fold degenerate Dirac point at the X point, as shown below.

With the symmetry under time reversal operation, $\Theta$, one can introduce another symmetry operation, $\widetilde{\Theta}$, by combining $\mathcal{S}$ with $\Theta$ such as $\widetilde{\Theta}\equiv\mathcal{S}\Theta$. Let us note that $\Theta$ commutes with all spatial transformations including $\mathcal{S}$, and its square is -1 for a spin-half particle. 
Since the square of $C_{2x}$ is also -1 for a spin-half particle, the square of $\widetilde{\Theta}$, then, becomes 
\[
  \widetilde{\Theta}^2=\mathcal{S}^2\Theta^2=\left(C_{2x}e^{ik_xa/2}\right)^2\Theta^2=e^{ik_xa}.
\]
It is important to note that $\widetilde{\Theta}^2$ becomes -1 at $k_x=\pi/a$, a zone boundary. From the theorem stating that any two-fold degeneracy must be protected by a symmetry that has an antiunitary operator with its square being -1~\cite{Hou2018}, the symmetry under $\widetilde{\Theta}$ guarantees Kramers degeneracy between $\ket{\psi}$ and $\ket{\widetilde{\Theta}\psi}$ if $\ket{\psi}$ is a Bloch eigenstate with $k_x=\pi/a$. Note that if $A$ is an antiunitary operator, then $\braket{f|g}=\braket{Af|Ag}^*=\braket{Ag|Af}$ for all vectors $\ket{f}$ and $\ket{g}$ in Hilbert space. Thus the antiunitary operator $\widetilde{\Theta}$ transforms
\[
  \braket{\psi|\widetilde{\Theta}\psi}
  =\braket{\widetilde{\Theta}\psi|\widetilde{\Theta}(\widetilde{\Theta}\psi)}^*=\braket{\widetilde{\Theta}^2\psi|\widetilde{\Theta}\psi}.
\]
Since $\widetilde{\Theta}^2=-1$ at $k_x=\pi/a$, we gets $\braket{\psi|\widetilde{\Theta}\psi}=-\braket{\psi|\widetilde{\Theta}\psi}$,
resulting in $\braket{\psi|\widetilde{\Theta}\psi}=0$. 
Since the invariant space under the operation of $\widetilde{\Theta}$ is $k_x=0$ and $k_x=\pi/a$, the band should be two-fold degenerate if $k_x=\pi/a$. It corresponds to the X$-$S line.

Together with the fact that there should be Weyl points on the $C_{2x}$-invariant $\Gamma-$X and S$-$Y lines~\cite{YoungKane2015}, the electronic structures with four bands should schematically look like Fig.~\ref{Fig3}(a). On the other hand $\mathcal{M}$ guarantees nodal lines that enclose the X and Y points~\cite{YoungKane2015}. As a result, the band lines should look like Fig.~\ref{Fig3}(b). The only way to let such two conditions satisfied simultaneously is to make the bands two-fold degenerate along the X$-$S line with the X point four-fold degenerate. 

We have also noted a certain condition in $\mathbf{k}$ under which $\mathcal{S}$ and $\mathcal{G}$ anticommute to each other as follows, which gives further degeneracy in the band structure. 
The anticommutator between $\mathcal{S}$ and $\mathcal{G}$ becomes
\begin{align*}
  \{\mathcal{S},\mathcal{G}\}
  &= i\sigma_xi\sigma_z\otimes R_x(\pi)e^{ik_xa/2}R_z(\pi)Pe^{i(k_xa+k_yb)/2}
  + i\sigma_zi\sigma_x\otimes R_z(\pi)Pe^{i(k_xa+k_yb)/2}R_x(\pi)e^{ik_xa/2}\\
  &=i\sigma_y\otimes \left(R_x(\pi)R_z(\pi)Pe^{ik_xa/2}e^{i(k_xa+k_yb)/2}
  - R_z(\pi)PR_x(\pi)e^{i(k_xa-k_yb)/2}e^{ik_xa/2}\right)\\
  &  =-2\sigma_y\otimes R_x(\pi)R_z(\pi)Pe^{ik_xa}\sin({k_yb/2}).
\end{align*}
Here, we have used $R_i(\pi)k_j=(-1)^{1-\delta_{ij}}k_jR_i(\pi)$ and $Pk_j=-k_jP$. It is then obvious that $\{\mathcal{S},\mathcal{G}\}=0$ if $k_y=0$, which guarantees two-fold degeneracy along the $\Gamma-$X line.~\bibnote{Let $\ket{k_x,+}$ an eigenstate of $\mathcal{S}$ with $k_y=0$ and an eigenvalue $+e^{ik_xa/2}$ which is a simultaneous eigenstate of the Hamiltonian; $\mathcal{S}\ket{+} = +e^{ik_xa/2}\ket{k_x,+}$. From the anticommutity of $\mathcal{S}$ and $\mathcal{G}$, $\mathcal{S}\mathcal{G}\ket{k_x,+} = -\mathcal{G}\mathcal{S}\ket{k_x,+} = -e^{ik_xa/2}\mathcal{G}\ket{k_x,+}$, which means that $\mathcal{G}\ket{k_x,+}$ is also an eigenstate of $\mathcal{S}$ with a different  eigenvalue $-e^{ik_xa/2}$. Since the symmetry of Hamiltonian under $\mathcal{G}$  guarantees that both $\ket{k_x,+}$ and $\mathcal{G}\ket{k_x,+}$ have the same energy,  it proves that the states with $k_y=0$ should be doubly degenerate.  It corresponds to the $\Gamma-$X line.}

Including all these factors, the band structure of a system symmetric under $\mathcal{S}$ and  $\mathcal{G}$ should be doubly degenerate along $\Gamma-$X$-$S line with a four-fold degeneracy at the X point~[Fig.~\ref{Fig3}(c)]. Such four-fold degenerate Dirac point is protected even with spin-orbit coupling properly taken into account and can be accessed with moderate doping for PO. Figure~\ref{Fig3}(d) shows the band structure of PO from our {\it ab initio} calculation, which contains all the essential features described in the schematic one shown in Fig.~\ref{Fig3}(c).
A final remark we would like to mention is that the same kind of Dirac point can be realized even at the Fermi level by replacing the oxygen atoms for PO$_{0.5}$ with atoms of odd number of valence electrons such as fluorene or sodium making such materials essentially metallic.


\section{Conclusions}
\label{Summary}

We present the structural, electronic, and vibrational properties  of phosphorene oxides PO$_x$ with $x\in[0,1]$ using \textit{ab initio} density  functional theory. Our calculated formation energy and energy gain show that the oxidation occurs naturally at least up to $x=1$. The electronic band gap increases with oxidation with $x$ for $x\lesssim0.4$ and then decreases. We have also found that two nonsymmorphic symmetries guarantee symmetry-protected degeneracies in the electronic band structures including a four-fold degenerate Dirac point accessible with moderate doping. The phonon dispersion relation shows that phosphorene oxides are structurally stable. We also propose that the degree of oxidation might be probed using IR analysis.

%% file: acknowledgements.tex
We acknowledge financial support from the Korean government through National Research Foundation (2019R1A2C1005417). Some portion of our computational work was done using the resources of the KISTI Supercomputing Center (KSC-2018-C2-0033 and KSC-2018-CHA0052).

%% file: My_collection.bbl
\providecommand{\latin}[1]{#1}
\makeatletter
\providecommand{\doi}
  {\begingroup\let\do\@makeother\dospecials
  \catcode`\{=1 \catcode`\}=2\doi@aux}
\providecommand{\doi@aux}[1]{\endgroup\texttt{#1}}
\makeatother
\providecommand*\mcitethebibliography{\thebibliography}
\csname @ifundefined\endcsname{endmcitethebibliography}
  {\let\endmcitethebibliography\endthebibliography}{}

%% file: nanolett.bbl
\begin{mcitethebibliography}{24}
\bibitem[Not()]{Note-1}
Let $\ket{k_x,+}$ an eigenstate of $\mathcal{S}$ with $k_y=0$ and an eigenvalue
  $+e^{ik_xa/2}$ which is a simultaneous eigenstate of the Hamiltonian;
  $\mathcal{S}\ket{+} = +e^{ik_xa/2}\ket{k_x,+}$. From the anticommutity of
  $\mathcal{S}$ and $\mathcal{G}$, $\mathcal{S}\mathcal{G}\ket{k_x,+} =
  -\mathcal{G}\mathcal{S}\ket{k_x,+} = -e^{ik_xa/2}\mathcal{G}\ket{k_x,+}$,
  which means that $\mathcal{G}\ket{k_x,+}$ is also an eigenstate of
  $\mathcal{S}$ with a different eigenvalue $-e^{ik_xa/2}$. Since the symmetry
  of Hamiltonian under $\mathcal{G}$ guarantees that both $\ket{k_x,+}$ and
  $\mathcal{G}\ket{k_x,+}$ have the same energy, it proves that the states with
  $k_y=0$ should be doubly degenerate. It corresponds to the $\Gamma-$X
  line.\relax
\mciteBstWouldAddEndPunctfalse
\mciteSetBstMidEndSepPunct{\mcitedefaultmidpunct}
{}{\mcitedefaultseppunct}\relax
\EndOfBibitem
\end{mcitethebibliography}
